\documentclass[prb,showpacs]{revtex4}

\usepackage{amsmath,amssymb}

\usepackage{graphicx}

\begin{document}

\widetext

\title{An estimate of the inter-system crossing time in light-emitting polymers}

\author{William Barford$^{1,2*}$ and Eric E. Moore$^{1**}$}

\affiliation{
$^1$Department of Physics and Astronomy, University
of Sheffield, Sheffield, S3 7RH, United Kingdom\\
$^2$Cavendish Laboratory, University of Cambridge,
Cambridge, CB3 0HE, United Kingdom
}

\begin{abstract}

The reported enhanced singlet-exciton yields in light-emitting polymers over the statistical limit of $25\%$ has attracted wide experimental and theoretical attention. Most theoretical estimates of the singlet-exciton yield depend crucially on estimates of the inter-system crossing rates induced by spin-orbit coupling. In this paper we use the experimentally determined phosphorescent life-time and energy of the lowest-lying triplet state, as well as calculated values of Huang-Rhys factors to estimate the spin-orbit matrix element and inter-system crossing time between the lowest-lying singlet and triplet states.

\end{abstract}


\maketitle

The reported enhanced singlet-exciton yields in light-emitting polymers over the statistical limit of $25\%$ has attracted wide experimental and theoretical attention. See the reviews \cite{17} and \cite{18} for references to the experimental and theoretical work. Most theoretical estimates of the singlet-exciton yield depend crucially on estimates of the inter-system crossing rates induced by spin-orbit coupling. A determination of this  rate would therefore be useful in attempting to resolve the issue of whether or not the singlet exciton yield is enhanced in light emitting polymers. In this paper we make an estimate of this rate.
We determine the spin-orbit matrix element between the lowest excited singlet and triplet states from the phosphorescence life-time of the triplet state. We use this matrix element to determine the inter-system crossing rate using the Fermi golden rule.

Emission
occurs from the `triplet' state, $T_1$, because it acquires some
singlet character as a result of the mixing with the `singlet'
state, $S_1$, induced by spin-orbit coupling, $H_{SO}$. Thus, the
experimentally determined lifetimes of $S_1$ and $T_1$ can be used
to estimate the matrix element of the spin-orbit coupling, $
\langle S_1 | H_{SO}|T_1\rangle$, and hence $\tau_{ISC}$. Assuming
that the spin-orbit coupling is much smaller than the exchange
energy, $\Delta$, we can employ perturbation theory.

We use the result from first order perturbation theory that the perturbed state, $|M'\rangle$, is,
\begin{equation}\label{}
    |M'\rangle = |M\rangle + \sum_{N \ne M} \frac{\langle N|H|M\rangle}{E_M - E_N} |N \rangle,
\end{equation}
where $H$ is the perturbation.
Here, the state $|N\rangle$ is a Born-Oppenheimer state, namely a direct product of an electronic state, $|n\rangle$, and a vibrational state associated with that electronic state, $|\nu;n\rangle$:
\begin{equation}\label{}
    |N\rangle = |n\rangle |\nu;n\rangle.
\end{equation} 

We require the corrected triplet state, $|T_1'\rangle$, resulting from the mixing with the singlet state, $|S_1\rangle$, via the spin-orbit coupling, $H_{SO}$:
\begin{eqnarray}\label{}
    |T_1'\rangle = |T_1\rangle + \sum_{\nu} \frac{\langle S_1|H_{SO}|T_1\rangle}{E_{T_1} - E_{S_1}} |S_1\rangle.
\end{eqnarray}
Here, the sum is over the vibrational levels of the singlet manifold.
In particular,  the zeroth vibronic state of the triplet manifold is
\begin{eqnarray}\label{Eq:4}
    |t'_1\rangle|0;{t'_1}\rangle &&= |t_1\rangle|0;{t_1}\rangle + 
   \sum_{\nu} \frac{\langle \nu;{s_1}|\langle s_1|H_{SO}|t_1\rangle|0;{t_1}\rangle}{E_{t_1}^0 - E_{s_1}^{\nu}} |s_1 \rangle|\nu;{s_1} \rangle
\\
\nonumber
&& = |t_1\rangle|0;{t_1}\rangle + 
  \langle s_1|H_{SO}|t_1\rangle \sum_{\nu} \frac{\langle \nu;{s_1}|0;{t_1}\rangle}{E_{t_1}^0 - E_{s_1}^{\nu}} |s_1 \rangle|\nu;{s_1}\rangle.
\end{eqnarray}

The general expression for the overlap of two vibrational wavefunctions is\cite{keil},
\begin{equation}
\langle\mu;i|\nu;j\rangle = \sqrt{\frac{\mu!}{\nu!}}\left(-\frac{a_{ij}}{\sqrt{2}}\right)^{(\nu-\mu)}\exp(-a_{ij}^2/4)L_{\mu}^{\nu-\mu}(a_{ij}^2/2),
\end{equation}
where $L_{m}^{n}(x)$ are the associated Laguerre polynomials,
\begin{equation}
L_m^n(x) = \sum_{k=0}^m\frac{(-1)^k(m+n)!}{(m-k)!(n+k)!k!}x^k,
\end{equation}
and
\begin{equation}\label{}
    a_{ij} = \sqrt{\frac{M\omega}{\hbar}}(Q_i-Q_j),
\end{equation}
is the difference in the dimensionless electron-lattice coupling
between the electronic states  $|i\rangle$ and $|j\rangle$.
The Huang-Rhys factor is,
\begin{equation}\label{}
    S_{ij} = \frac{a_{ij}^2}{2}.
\end{equation}
Thus, the vibrational overlap between the zeroth level of the triplet manifold and the $\nu$th level of the singlet manifold is,
\begin{equation}\label{}
    \langle \nu;{s_1}|0;{t_1}\rangle =  \exp(-\textsf{S}/2)\sqrt{\frac{\textsf{S}^{\nu}}{\nu!}}\left[\textrm{sgn}(Q_t-Q_s)\right]^{\nu},
\end{equation}
where $\textsf{S}$ is the Huang-Rhys factor for the singlet relative to the triplet and $\textrm{sgn}(Q_t-Q_s)$ is the relative sign of the difference in the configurational coordinates of the singlet and triplet states. 

We therefore write Eq (\ref{Eq:4}) as,
\begin{eqnarray}\label{Eq:6}
 |t'_1\rangle|0;{t'_1}\rangle = |t_1\rangle|0;{t_1}\rangle - 
  W  |s_1 \rangle\sum_{\nu} A_{\nu}|\nu;{s_1}\rangle,
\end{eqnarray}
where,
\begin{equation}\label{}
    A_{\nu} = \frac{\exp(-\textsf{S}/2)^{\nu/2}}{\sqrt{\nu!}(\Delta + \nu \hbar\omega)}\left[\textrm{sgn}(Q_t-Q_s)\right]^{\nu},
\end{equation}
  $\Delta$ is the $0-0$ energy difference between the singlet and triplet states, $\hbar\omega$ is the phonon frequency and 
\begin{equation}\label{}
    W = \langle s_1|H_{SO}|t_1\rangle, 
\end{equation}
is the electronic spin-orbit matrix element, which we wish to determine.

The radiative life-times are derived from the Einstein expression,
\begin{equation}\label{Eq:8}
    \tau^{-1}_M = \left( \frac{\textrm{e}^2}{3\pi\epsilon_0\hbar^4 c^3}\right) 
    \Delta E_M^3 \langle S_0|\hat{\mu}|M \rangle ^2,
\end{equation}
where $\Delta E_M$ is the transition energy and $|S_0\rangle$ is the ground state.

We  consider the life-time of the vertical transition from the zeroth vibronic level of the triplet state to the $\nu_{t_1}th$ level of the ground state, where $\nu_{t_1}$ is defined by
\begin{equation}\label{}
    \nu_{t_1} = S_{t_1} - 1/2
\end{equation}
and $S_{t_1}$ is the Huang-Rhys factor of the triplet state relative to the ground state.
Calculating the matrix element $\langle \nu_{t_1};s_0|\langle s_0|\hat{\mu} |t'_1 \rangle |0;{t'_1}\rangle $ using Eq.\ (\ref{Eq:6}) gives,
\begin{eqnarray}\label{Eq:15}
&& \langle \nu_{t_1};s_0|\langle s_0|\hat{\mu} |t_1' \rangle |0;{t'_1}\rangle = 
W  \langle s_0|\hat{\mu}|s_1 \rangle\sum_{\nu} A_{\nu}\langle \nu_{t_1};s_0|\nu;{s_1}\rangle
\nonumber
\\
&& = 
W  \langle s_0|\hat{\mu}|s_1 \rangle\langle \nu_{s_1};s_0|0;s_1\rangle
\sum_{\nu} \frac{A_{\nu}\langle \nu_{t_1};s_0|\nu;{s_1}\rangle}{ \langle\nu_{s_1};s_0|0;s_1\rangle}
\end{eqnarray}
as $\langle s_0|\hat{\mu}|t_1\rangle = 0$. $\nu_{s_1}= S_{s_1} - 1/2$, where $S_{s_1}$ is the Huang-Rhys factor for the singlet state relative to the ground state.

Now,
\begin{equation}
 \langle \nu_{t_1};s_0|\langle s_0|\hat{\mu} |t_1' \rangle |0;{t'_1}\rangle = 
\langle s_0|\hat{\mu} |t_1' \rangle  \langle \nu_{t_1};s_0|0;{t'_1}\rangle
=
\langle S_0|\hat{\mu} |T_1'\rangle^{v}
\end{equation} 
and 
\begin{equation}
 \langle \nu_{s_1};s_0|\langle s_0|\hat{\mu} |s_1 \rangle |0;{s_1}\rangle = 
\langle s_0|\hat{\mu} |s_1 \rangle \langle \nu_{s_1};s_0|0;{s_1}\rangle =
\langle S_0|\hat{\mu} |S_1\rangle^{v}
\end{equation}
for the vertical transitions from the triplet and singlet states, respectively.

Thus, Eq (\ref{Eq:15}) becomes,
\begin{equation}\label{}
 \langle S_0|\hat{\mu} |T_1'\rangle^{v} =  \langle S_0|\hat{\mu} |S_1\rangle^{v}W X,
\end{equation}
where we define,
\begin{equation}\label{Eq:19}
    X = \sum_{\nu} \frac{A_{\nu}\langle \nu_{t_1};s_0|\nu;{s_1}\rangle}{ \langle\nu_{s_1};s_0|0;s_1\rangle}.
\end{equation}

The expression for the ratio of the life-times is therefore
\begin{equation}\label{Eq:12}
    \frac{\tau_{S_1}}{\tau_{T_1}} = \left(\frac{\Delta E_{T_1}}{\Delta
    E_{S_1}}\right)^3  |W|^2 X^2,
\end{equation}
where $\Delta E_{Y_1}$ is the  excitation energy of of the state $Y_1$.

The Huang-Rhys factors required for the determination of $X$ (Eq (\ref{Eq:19})) are estimated from density matrix renormalization group calculations of the Pariser-Parr-Pople-Peierls model for poly(p-phenylene) oligomers\cite{moore}. We find for short oligomers ($4 -8$ phenyl rings) that $\textsf{S} \sim 1.1$, $S_{t_1} \sim 2.9$ and $S_{s_1} \sim 0.9$.
In PPV-DOO the radiative lifetimes of $S_1$ is $200$ ps
\cite{21}, while $\Delta
E_{S_1} = 2.6$ eV \cite{21}, $\Delta E_{T_1} = 1.5$eV \cite{22},
and thus $\Delta = 1.1$ eV. In PPP derivatives the radiative life time of $T_1$ is of order one second\cite{bassler}.

We note that each term in the sum over $\nu$ in the definition of $X$ contains the factor,
\begin{equation}\label{}
   \left[ \textrm{sgn}(Q_{t_1}-Q_{s_1})\textrm{sgn}(Q_{s_1}-Q_{s_0})\right]^{\nu} \equiv b^{\nu}.
\end{equation}
The value of $b$ ($=\pm 1$) determines the value of $X$. 
If $b=+1$, $X = 1.15$ eV$^{-1}$, whereas if $b=-1$, $X = -1.27$ eV$^{-1}$. Since only the absolute magnitude is important, and because the sign of $b$ is unknown we take $|X|= 1.21$ eV$^{-1}$.

Using these values in Eq.\ (\ref{Eq:12}) gives
$|W| \sim 5 \times 10^{-5} $ eV.

Inter-system crossing from $S_1$ to $T_1$ is an iso-energetic transition from the
lowest vibrational level of $S_1$ to the vibrational level, $\nu$,
of $T_1$ \textit{at the same energy}. The  rate, $k^{ISC}$, induced by
spin-orbit coupling is given by the Fermi Golden Rule expression (see, for example ref\cite{robinson}):
\begin{equation}\label{}
    k^{ISC} = \frac{2\pi}{\hbar} |W|^2 F_{0\nu} \rho_f(E),
\end{equation}
where $F_{0\nu}$ is the Franck-Condon factor, $\langle0|\nu\rangle^2$ and $\rho_f(E)$ is the final density of states, which we take to be
the inverse of the phonon energy, \emph{i.e.}\ $5$ eV$^{-1}$.

Thus,
\begin{equation}\label{}
    k^{ISC} = \frac{2\pi}{\hbar} |W|^2 \exp(-\textsf{S})
    \frac{\textsf{S}^{\nu}}{\nu!}\rho_f(E).
\end{equation}
Now, $\nu = 1.1/0.2 = 5.5$ and 
taking $\textsf{S}=1.1$ one finds that spin-orbit coupling $\tau_{ISC}$ is $5$ $\mu$s. This rate is very low, and is not compatible with a triplet yield under photo-excitation of a few percent.

Notice that inter-system crossing between $S_1$ and $T_1$ is inhibited by the exponentially small Franck-Condon factor arising from the large exchange energy between these states. However, inter-system crossing between  quasi-degenerate  states is not inhibited by this factor. It is highly likely that there is a higher lying pseudo-momentum counterpart to $T_1$, $T_n$ say, close in energy to $S_1$ (see, for example refs \cite{barford2} and \cite{burin}). Inter-system crossing to $T_n$ would rapidly yield $T_1$ excitons via efficient inter-conversion mechanisms.
Assuming that $S_1$ and $T_n$ are quasi-degenerate the inter-system crossing time between these states is approximately $500$ times smaller than that between $S_1$ and $T_1$, namely $\sim 10$ ns. Such a time is compatible with a triplet yield of a few percent. This value is also consistent with an inter-system crossing time of $4$ ns quoted in ref\cite{frolov}.

In conclusion, we have estimated the intersystem crossing time between quasi-degenerate states in the lowest singlet and triplet manifolds to be $\sim 10$ ns. Some theories assume that intersystem crossing between the higher lying quasi-degenerate charge transfer excitons is responsible for the enhanced singlet exciton yield (see, for example ref\cite{barford}). The calculation presented here does not directly address this intersystem crossing time, however it is perhaps reasonable to assume that both rates are comparable.


\begin{acknowledgements}
This work was supported by the EPSRC (U.K.) (GR/R02177/01 and GR/R03921/01).
W. B. thanks R. H. Friend, N. C. Greenham and A. K\"ohler for useful discussions.
W.B. also gratefully acknowledges the financial support of the Leverhulme
Trust, and thanks the Cavendish Laboratory and Clare Hall,
Cambridge for their hospitality.
\end{acknowledgements}


\begin{references}

\bibitem[]{email}
E.mail address: W.Barford@sheffield.ac.uk

\bibitem[] {address1} 
$^*$Permanent address: Department of Physics and Astronomy, University of Sheffield.
\bibitem[] {address2}
$^{**}$Current address: 392 Central Park West, New York, NY 10025, U.S.A. 

\bibitem{17} M. Wohlgennant and Z. V. Vardeny, \textit{J. Phys.: Condens. Matter} \textbf{15}, R83-R107 (2003).

\bibitem{18} A. K\"ohler and J.  Wilson, \textit{Organic Electronics}, \textbf{4}, 179 (2003).

\bibitem{keil} T. Keil, \textit{Phys. Rev.} \textbf{140}, A601 (1965).


\bibitem{moore} E. E. Moore, W. Barford and R. J. Bursill, submitted to Phys Rev B..

\bibitem{21} S. V. Frolov, Z. Bao, M. Wohlgennant, and Z. V. Vardeny,
\emph{Phys. Rev.} \textbf{B65}, 205209 (2002).

\bibitem{22} A. P. Monkman, H. D. Burrows, L. J. Hartwell, L. E. Horsburg,
I. Hamblett, and S. Navaratnam, \emph{Phys. Rev. Lett.}
\textbf{86}, 1358-1361 (2001).

\bibitem{bassler} D. Hertzel, Y. V. Romanovskii, B. Schweitzer, U. Sherf and H. B\"assler, Macromolecular Symposia, \textbf{175}, 14 (2001).

\bibitem{robinson} G. W. Robinson and R. P. Frosch, \textit{J. Chem. Phys.} \textbf{37}, 1962 (1962),
G. W. Robinson and R. P. Frosch, \textit{J. Chem. Phys.} \textbf{38}, 1187 (1963).

\bibitem{barford2} W. Barford, R. J. Bursill and R. W. Smith, \textit{Phys. Rev.} \textbf{B66}, 115205 (2002).

\bibitem{burin} A. L. Burin and M. A. Ratner, \textit{J. Chem. Phys.} \textbf{109}, 6092 (1998).

\bibitem{frolov} S. V. Frolov, M. Liess, P. A. Lane, W. Gellermann, and Z. Vardeny, \textit{Phys. Rev. Lett.} \textbf{78}, 4285 (1997).

\bibitem{barford} W. Barford, \textit{Phys. Rev.} \textbf{B70} (in press).


\end{references}
\end{document}